\documentclass[acus]{JAC2000}

\usepackage{graphicx}

\newcommand{\grad}{\mbox{$^{\circ}$}}
\newcommand{\degC}{\mbox{\grad{C~}}}

\begin{document}
\title{\flushright{TUAP050}\\[15pt] \centering CORRECTOR POWER SUPPLIES
  WITH A DAC RESOLUTION UP TO 24 BITS BASED ON 16 BIT DAC
  DEVICES\thanks{Funded by the Bundesministerium f\"ur Bildung,
  Wissenschaft, Forschung und Technologie (BMBF) and the Land Berlin}} 

\author{ K.~B\"urkmann, B.~Franksen, R.~Lange, I.~M\"uller, 
         R.~M\"uller, J.~Rahn, T.~Schneegans, BESSY, Berlin, FRG \\
         G.~v.~Egan, EuKontroll, Berlin, FRG }

\maketitle

\begin{abstract}
At BESSY the standard 16 bit resolution of the corrector power
supplies was insufficient for the continuous orbit drift correction
\cite{orbit-control}. A new sophisticated design of the analog/digital
I/O board \cite{can-china} increases the resolution of the analog
output up to 24 bits without suffering on losses in the long term
stability or the dynamic range of the correctors. This is achieved by
a cost-efficient board design using standard 16 bit DAC devices in a
range-overlapping fine/coarse architecture.
\end{abstract}

\section{INTRODUCTION}

The third generation light source \mbox{BESSY~II} started operation
equipped with power supplies using low thermal drift 16 Bit DAC
devices. First experiences with the implemented SVD based automatic
orbit correction scheme showed that the 16 bit resolution of the power
supplies was not sufficient to correct the orbit without unacceptable
influences to specific experiments\cite{orbit-control}. \\
A typical solution to solve this problem is to reduce the dynamic
range of the correctors. A single bit of the DAC device then
represents a smaller current step of the corrector power supply
output. The obvious disadvantage of this solution is that large kicks
and bumps are not achievable. This leads to an inacceptably restricted
use of the power supplies for diagnostic and recalibration purposes.\\
To avoid the disadvantages of a reduced dynamic range a new I/O
board with an increased DAC channel resolution of up to 24 bits
has been introduced.

\section{24 BIT DAC BOARD DESIGN}

\subsection{Compatibility Demands}

At BESSY the power supplies are CAN bus controllable. Therefore an I/O
board, together with a piggy back embedded controller including the
CAN bus interface, is plugged directly into the power supply. This
board combination provides the whole functionality needed for low
level control of a typical power supply\cite{can-china}.\\ \newpage
Taking this into account, a new 24 bit I/O board, ADA2x16-IO8, has been
developed\footnote{The design, development and production of this
board has been performed by the EuKontroll GmbH, Berlin, Germany. For
detailed information please contact Georg~v.~Egan,
EuKontroll@t-online.de}.
The design has the same form factor as the former
16 bit I/O board. The new board is fully compatible to the former
design and consists of:

\begin{itemize}
\item 8 digital inputs and 8 digital outputs
\item a 24 bit analog output
\item a fast 16 bit flash ADC, multiplexed to 4 inputs
\item a slow dual slope $\pm$~15~bit~$+$~sign ADC, multiplexed to 4 inputs 
\item a connector to house the BESSY embedded controller including a
      CAN bus interface 
\item a configurable bus interface unit supporting several bus types
      (e.g. ISA96, VME) 
\end{itemize}

\subsection{Analog Output Stage Design}

The long term stability of the corrector power supplies directly
affects the static orbit stability of the BESSY~II storage ring. This
requires a high stability analog stage with low thermal drifts (in our
case typically 1.5~ppm~/~\degC or better) for the I/O board. The long
term stability depends mainly on the drift of the voltage reference
and the DAC output. The significant differences in thermal drifts of
typical DACs available on the market are shown in
Table~\ref{tab:dac-comp}.

\begin{table}[htbp]
  \begin{center}
\begin{small}
    \begin{tabular}[c]{|l|l|l|}
      \hline 
      & PCM1704 & AD7846 \\
      & Audio DAC & Monolithic DAC \\
      & 24 bit & 16 bit \\ \hline
      Gain Error   & $\pm$ 3~\%~FSR & $\pm$~0.05~\%~FSR \\
      Bipolar Zero & $\pm$ 1~\%~FSR & $\pm$~0.024~\%~FSR \\
      Error        &              & \\
      Gain TC      & $\pm$~25~ppm~FSR~/~\degC & $\pm$~1~ppm~FSR~/~\degC \\
      Zero TC      & $\pm$~5~ppm~FSR~/~\degC & $\pm$~1~ppm~FSR~/~\degC \\ \hline
       \multicolumn{3}{|l|}{FSR = full scale range, TC = temperature coefficient}\\
      \hline
    \end{tabular}
\end{small}
  \end{center}
  \caption{Errors and Drifts of Typical 24 Bit Audio DAC vs. 16 Bit Monolithic DAC}
  \label{tab:dac-comp}
\end{table}

\begin{figure}[htbp]
  \begin{center}
  \includegraphics*[width=80.0mm]{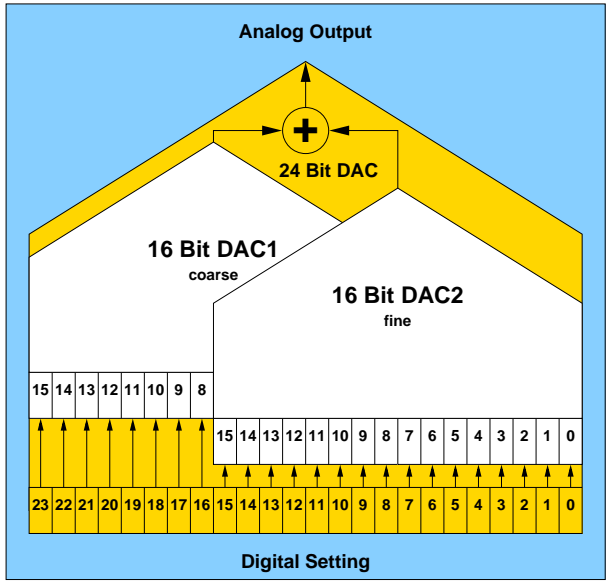}
  \caption{24 Bit DAC Stage Principle}
  \label{fig:dac-stage}
  \end{center}
\end{figure}

Because long term stability of the power supplies is a key feature
to achieve a sufficient static orbit stability, the design of the 24
bit DAC board is based on a combination of two low thermal drift 16
bit DAC devices. A range-overlapping architecture is used to
accomplish the 24 bit resolution (see Figure \ref{fig:dac-stage}). The
16 bit DAC devices are the same as in the design of the 16 bit
board.  The higher 8 bits of DAC1 are used for coarse setting and
the 16 bits of DAC2 for fine setting. The lower 8 bits of DAC1 are
available to linearize the relative accuracy or endpoint nonlinearity
of the 24 bit output if needed (in our case this feature is not used).

Critical in this design is the transition point between the two 16 bit
DAC devices (e.g. bit 16 switches to one and bits 0..15 are switching
to zero). This directly affects the differential nonlinearity and
therefore the monotonicity of the design. In our case the board is
calibrated to provide monotonicity up to 17 bits resolution. Considering
a full scale range of $\pm$~10~V a single bit of 24 bit resolution
represents 1.192~$\mu$V and 17 bit represents 152.59~$\mu$V; i.e. the
design is calibrated to be monotonic for relative settings down to
152.59~$\mu$V or better.

\subsection{Measurements}

\begin{figure}[htbp]
  \begin{center}
  \includegraphics*[width=80.0mm]{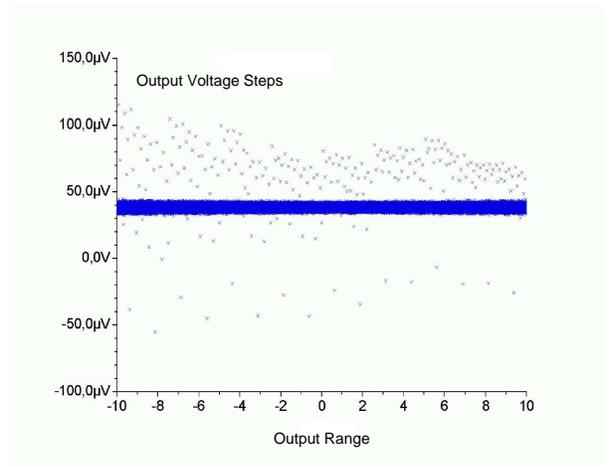}
  \caption{Output Voltage Steps vs. Input Setting Steps (19 Bit
Operation Mode)}
  \label{fig:delta-ist}
  \end{center}
\end{figure}

\begin{figure}[htbp]
  \begin{center}
  \includegraphics*[width=80.0mm]{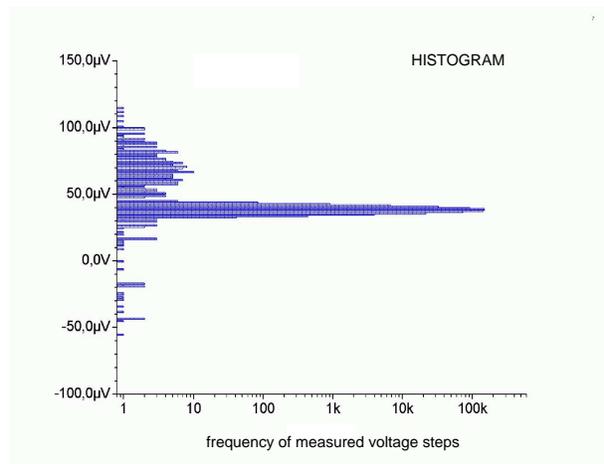}
  \caption{Histogram of Output Voltage Steps (19 Bit Operation Mode)}
  \label{fig:delta-ist-histo}
  \end{center}
\end{figure}

Figure \ref{fig:delta-ist} shows the measured relative output steps of
one randomly selected 24 bit DAC when the input is incremented by
steps of 2$^5$ digits, which is equivalent to a 19 bit operation
mode. In the ideal case of a 19 bit operation mode, every relative
output step of the DAC should be 38.15~$\mu$V. In Figure
\ref{fig:delta-ist-histo} the majority of measured
output steps are in the 38.15~$\mu$V region and only some few are in
the 152.59~$\mu$V region.\\ 
Because of these encouraging measurement results we expected significant
improvements regarding the resolution of the corrector power supplies
and therefore performance improvements of the orbit correction
scheme.

\section{PERFORMANCE}

\begin{figure}[htbp]
  \begin{center}
  \includegraphics*[width=80.0mm]{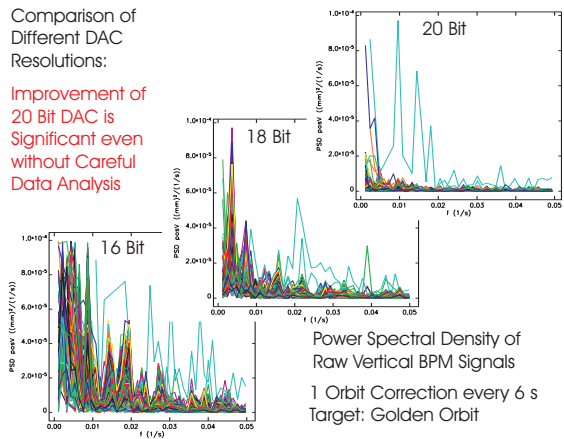}
  \caption{FFT of Time Dependent Beam Position Data}
  \label{fig:SVD-performance}
  \end{center}
\end{figure}

After installation of the ADA2x16-IO8 24 bit DAC boards, first
tests with the SVD based orbit correction scheme have been done using
different resolutions for the setting. Figure
\ref{fig:SVD-performance} shows the FFT performed over the resulting
time dependent vertical beam position data of all beam position
monitors (BPMs). A drastically reduced orbit drift could be seen when
the resolution was increased from 16 bit to 18 resp. 20 bit
\cite{pac01}.

\section{CONCLUSION}

The new 24 bit design of the analog/digital I/O board provides the
higher resolution needed for the correctors of the third generation
light source \mbox{BESSY~II}. Due to the range-overlapping
architecture of the board using the low thermal drift 16 bit DAC
devices, the known static orbit stability is guaranteed even without
active orbit correction. Applying a SVD based orbit drift correction
algorithm the closed orbit is now being corrected with a stability of
typically \mbox{$<$~1..2$\mu$m}.

\end{document}